 \documentclass[preprint,onecolumn,amsmath,amssymb,aps,prb]{revtex4-1}

\usepackage{graphicx}% Include figure files
\usepackage{dcolumn}% Align table columns on decimal point
\usepackage{bm}% bold math
 \usepackage{float} 
\usepackage{fullpage}
\usepackage{color}
\usepackage{subcaption}
\usepackage{multirow}
\usepackage{amsmath}
\usepackage{amssymb}
\usepackage{amsthm}
\usepackage[colorlinks   = true, urlcolor     = cyan, linkcolor    = blue, citecolor   = red]{hyperref}
\usepackage{bm}
\usepackage{pict2e}
\usepackage{epstopdf}
\usepackage{comment}
\usepackage{epsfig}
\usepackage{url}
\usepackage[ruled,vlined]{algorithm2e}
\newcommand{\beq}{\begin{equation}}
\newcommand{\eeq}{\end{equation}}
\newcommand{\beqs}{\begin{eqnarray}}
\newcommand{\eeqs}{\end{eqnarray}}
\newcommand{\beql}{\begin{equation} \label}

\usepackage{cancel}
\usepackage{amsmath}
\usepackage{amssymb}
\usepackage{etoolbox}
\usepackage{graphicx}
 \usepackage[english]{babel}
\usepackage{mathrsfs} % Bilinear forms
\usepackage{enumitem}                                  % Fancy enumerate, e.g. a) b) c)

\usepackage{amsthm} % Theorems
\usepackage{siunitx}
\usepackage{ upgreek }
\begin{document}

%\preprint{APS/123-QED}

\title{Torsional moduli of transition metal dichalcogenide nanotubes from first principles}

\author{Arpit Bhardwaj}
\affiliation{College of Engineering, Georgia Institute of Technology, Atlanta, GA 30332, USA}

\author{Abhiraj Sharma}
\affiliation{College of Engineering, Georgia Institute of Technology, Atlanta, GA 30332, USA}

\author{Phanish Suryanarayana}
\email{phanish.suryanarayana@ce.gatech.edu}
%\homepage{https://www.phanishgroup.com}
%\email{phanish.suryanarayana@ce.gatech.edu}
\affiliation{College of Engineering, Georgia Institute of Technology, Atlanta, GA 30332, USA}

%\date{\today}

\begin{abstract}
We calculate the torsional moduli of  single-walled transition metal dichalcogenide (TMD) nanotubes using \emph{ab initio} density functional theory (DFT). Specifically, considering forty-five select TMD nanotubes, we perform symmetry-adapted DFT calculations to calculate the torsional moduli for the armchair and zigzag variants of these materials in the low-twist regime and at practically relevant diameters. We find that the torsional moduli follow the trend: MS\textsubscript{2} $>$ MSe\textsubscript{2} $>$ MTe\textsubscript{2}. In addition, the moduli display a power law dependence on diameter, with the scaling generally close to cubic, as predicted by the isotropic elastic continuum model.  In particular, the shear moduli so computed are in good agreement with those predicted by the isotropic relation in terms of the Young's modulus and Poisson's ratio, both of which are also calculated  using symmetry-adapted DFT.  Finally, we develop a linear regression model for the torsional moduli of TMD nanotubes based on the nature/characteristics of the metal-chalcogen bond, and  show that it is capable of making reasonably accurate predictions.  
\end{abstract}

\keywords{Torsional modulus, Transition Metal Dichalcogenides, Nanotubes, Density Functional Theory, Shear modulus, Young's modulus, Poisson's ratio}

\maketitle

\section{Introduction} The synthesis of carbon nanotubes around three decades ago \cite{iijima1991helical} has revolutionalized the fields of nanoscience and nanotechnology. Even in the specific instance of nanotubes --- quasi one-dimensional hollow cylindrical structures with diameters in the nanometer range --- nearly two dozen nanotubes have now been synthesized \cite{tenne2003advances, rao2003inorganic, serra2019overview}, with the potential for thousands more given the large number of stable two-dimensional materials  that have been predicted from first principles  calculations \cite{haastrup2018computational, zhou20192dmatpedia}. Nanotubes have been the subject of intensive research, inspired by the novel and enhanced mechanical, electronic, optical, and thermal properties relative to their bulk counterparts \cite{tenne2003advances, rao2003inorganic, serra2019overview}.

Nanotubes can be categorized based on the classification adopted for the corresponding two-dimensional materials from which they can be thought to be constructed. Among  the different groups, the transition metal dichalcogenide (TMD) group --- materials of the form MX\textsubscript{2}, where M and X represent a transition metal and chalcogen, respectively --- is currently the most diverse, particularly given that they contain the largest number of distinct  nanotubes  synthesized to date \cite{tenne2003advances, rao2003inorganic, serra2019overview}. TMD nanotubes have a number of interesting properties including high tensile strength \cite{kis2003shear, kaplan2007mechanical, kaplan2006mechanical, tang2013revealing}, mechanically tunable electronic properties \cite{zibouche2014electromechanical, oshima2020geometrical, li2014strain, ghorbani2013electromechanics, lu2012strain,ansari2015ab, levi2015nanotube}, and low cytotoxicity \cite{pardo2014low}.  These properties make TMD nanotubes suited to a number of   applications, including reinforcement of composites \cite{shtein2013fracture, otorgust2017important, simic2019impact, nadiv2016critical, huang2016advanced, naffakh2016polymer}, nanoelectromechanical (NEMS) devices \cite{yudilevichself, levi2015nanotube, divon2017torsional}, and medicine \cite{lalwani2013tungsten}, where knowledge of their mechanical properties is important from the perspective of  both design and performance.   

In view of the above, there have been a number of efforts to characterize the elastic properties of  TMD nanotubes, both experimentally \cite{kaplan2004mechanical, kaplan2006mechanical, wang2008situ,grillo2020ws2} and theoretically \cite{zibouche2014electromechanical, wang2016strain, bandura2018calculation, xiong2017effects, ansari2015ab, ying2020mechanical, xiao2014theoretical, lorenz2012theoretical, li2014strain, sorkin2014nanoscale, kalfon2012insights}. However,  these studies are limited to only a few TMDs, and that too only for the case of axial tension/compression. In particular, determining the torsional moduli for these systems --- relevant for applications such  as resonators in NEMS devices \cite{divon2017torsional} --- has been limited to very few experimental \cite{nagapriya2008torsional, divon2017torsional} and theoretical \cite{zhang2010helical,bucholz2012mechanical} research works, and that too only for a couple of materials.   Indeed, the study of torsional deformations at practically relevant twists and nanotube diameters is intractable to ab initio methods like Kohn-Sham density functional theory (DFT) \cite{hohenberg1964inhomogeneous, kohn1965self} --- expected to provide higher fidelity than tight binding  and force field calculations for nanoscale systems --- given the large number of atoms that are required when employing the standard periodic boundary conditions \cite{sharma2021real}.  Therefore, accurate estimates for a fundamental mechanical property like torsional modulus is not available for TMD nanotubes, which provides the motivation for the current work. 

In this work, we calculate the torsional moduli of  forty-five select single-walled armchair and zigzag TMD  nanotubes using Kohn-Sham DFT. Specifically, considering nanotubes that have been synthesized or are expected  to be so in the future, we perform symmetry-adapted DFT calculations to calculate the torsional moduli of these materials at practically relevant  twists and nanotube diameters. We find the following relation for the torsional moduli values: MS\textsubscript{2} $>$ MSe\textsubscript{2} $>$ MTe\textsubscript{2}.  In addition, we find that the moduli display a power law dependence on diameter, with a scaling that is generally close to cubic, as predicted by the isotropic elastic continuum model.  In particular,  the shear moduli so determined are in good agreement with that predicted by the isotropic relation in terms of the Young's modulus and Poisson's ratio, both of which are also calculated  in this work using symmetry-adapted DFT.  We also develop a linear regression model for the torsional moduli of TMD nanotubes based on the nature and characteristics of the metal-chalcogen bond, and  show that it is capable of making reasonably accurate predictions. 

The remainder of the manuscript is organized as follows. In Section~\ref{Sec:Methods}, we discuss the chosen TMD nanotubes and describe the symmetry-adapted DFT simulations for calculation of their torsional moduli.  Next, we  present and discuss the results obtained  in Section~\ref{Sec:Results}. Finally, we provide concluding remarks in Section~\ref{Sec:Conclusions}

%%%%%%%%%%%%%%%%%%%%%%%%%%%%%%%%%%
%%%%%%%%%%%%%%%%%%%%%%%%%%%%%%%%%%%%%%%%%%%%

\section{Systems and methods} \label{Sec:Methods} We consider the following single-walled TMD nanotubes with 2H-t symmetry \cite{nath2002nanotubes, bandura2014tis2}: M$=$\{V, Nb, Ta, Cr, Mo, W, Fe, Cu\} and X$=$\{S, Se, Te\}; and the following ones with 1T-o symmetry \cite{nath2002nanotubes, bandura2014tis2}: M$=$\{Ti, Zr, Hf, Mn, Ni, Pd, Pt\} and X$=$\{S, Se, Te\}. These materials have been selected among all the possible transition metal-chalcogen combinations as they have either been synthesized as single/multi-walled nanotubes \cite{nath2001mose2, nath2001simple, chen2003titanium, nath2001new,nath2002nanotubes, tenne1992polyhedral, gordon2008singular, bruser2014single, remskar2001self} or the corresponding two-dimensional atomic monolayers have been predicted to be stable from ab initio calculations \cite{haastrup2018computational, heine2015transition, guo2014tuning}. The radii for  these nanotubes have been chosen so as to be commensurate with those  that have been experimentally synthesized, and in cases where such data is not available, we choose radii commensurate with synthesized nanotubes that are expected to have similar structure. 

\begin{figure}[htbp!]
\centering
\includegraphics[width=0.63\textwidth]{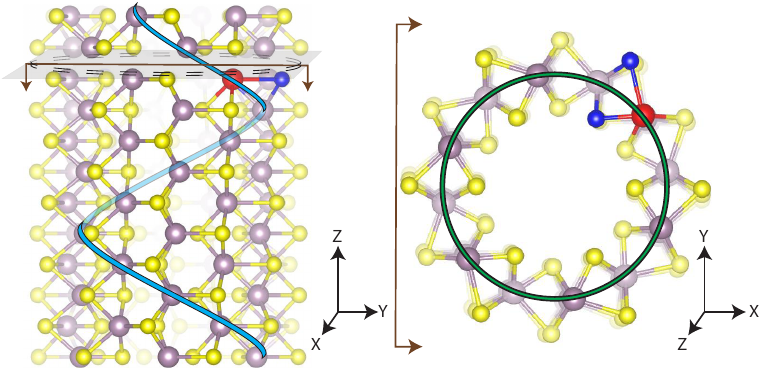}
\caption{Illustration showing the cyclic and helical symmetry present in a twisted (6,6) TMD nanotube with 2H-t symmetry. In particular, all atoms in the nanotube can be considered to be cyclic and/or helical images of the metal and chalcogen atoms that have been colored red and blue, respectively. This symmetry is exploited while performing electronic structure simulations using the Cyclix-DFT code \cite{sharma2021real}.}
\label{fig:illustration}
\end{figure}

We utilize the Cyclix-DFT code \cite{sharma2021real} --- adaptation of the state-of-the-art real-space DFT code SPARC \cite{xu2020sparc, ghosh2017sparc1, ghosh2017sparc2} to cylindrical and helical coordinate systems, with the ability to  exploit cyclic and helical symmetry in one-dimensional nanostructures \cite{sharma2021real, ghosh2019symmetry, banerjee2016cyclic} --- to calculate the torsional moduli of the aforementioned TMD nanotubes in the low twist limit. Specifically, we consider three-atom unit cell/fundamental domains that has one metal atom and two chalcogen atoms, as illustrated in Fig.~\ref{fig:illustration}. Indeed, such calculations are impractical without the symmetry adaption, e.g., a (57,57) MoS\textsubscript{2}  nanotube (diameter $\sim 10$ nm) with an external twist of $2\times10\textsuperscript{-4}$ rad/Bohr has $234,783$ atoms in the simulation domain when employing periodic boundary conditions, well beyond the reach of even state-of-the-art DFT codes  on large-scale parallel machines \cite{banerjee2018two, motamarri2020dft, xu2020sparc}. It is worth noting that the Cyclix-DFT code has already been successfully employed for the study of physical applications \cite{PhysRevMaterials.5.L030801, kumar2021flexoelectricity, sharma2021real, kumar2020bending}, which provides evidence of its accuracy.

We employ optimized norm-conserving Vanderbilt (ONCV) \cite{hamann2013optimized} pseudopotentials from the SG15 \cite{SCHLIPF201536} collection and the semilocal Perdew–Burke–Ernzerhof (PBE) \cite{perdew1986accurate} exchange-correlation functional. Apart from the tests by the developers \cite{SCHLIPF201536}, we have verified the transferability of the chosen pseudopotentials by comparisons with  all-electron DFT code Elk \cite{elk} for select bulk systems. In addition, we have found that the equilibrium geometries of the nanotubes and their two-dimensional counterparts (Supplementary Material) are in very good agreement with previous DFT results  \cite{haastrup2018computational, zhou20192dmatpedia, wang2016strain, xiao2014theoretical, li2014strain,ataca2012stable, chang2013orbital, amin2014strain, guo2014tuning}. There is also very good agreement with experimental measurements \cite{chen2003titanium, nath2003superconducting, nath2002nanotubes, nath2001mose2}, confirming the suitability of the chosen exchange-correlation functional.  Since we are interested in torsional moduli for the low-twist regime --- corresponds to small (linear) perturbations of electron density from the undeformed nanotube --- the use of more sophisticated functionals and/or inclusion of relativistic effects through spin orbit coupling (SOC) are not expected to  change the results noticeably, especially considering that significant error cancellations occur while taking differences in energy.  This is evidenced by the small differences in the ground state electron density between PBE and more sophisticated hybrid functionals for the TMD monolayer systems, even in the presence of SOC \cite{kumar2020bending}.

We calculate the torsional modulus in the low-twist regime by first performing ground state DFT simulations for various twisted configurations of the nanotube, and then fitting the data to the following quadratic relation: 
\begin{equation}
\mathcal{E} (d, \theta) = \mathcal{E} (d, 0) + \frac{1}{2} K(d) \theta^2 \,,
\end{equation}
where $K$ is the torsional modulus, $d$ is the diameter of the nanotube, and $\theta$ and $\mathcal{E}$ are the twist and ground state energy densities, respectively, i.e., defined per unit length of the nanotube. Indeed, small enough twists are chosen so that linear response is observed, i.e., the torsional modulus is independent of the the twist (Supplementary Material). It is important to note that the resulting shear strains --- quantity that better describes the behavior/response of nanotubes, by allowing systematic comparison between tubes with different diameters --- are commensurate with those found in torsion experiments \cite{levi2015nanotube, divon2017torsional, nagapriya2008torsional}. All numerical parameters in Cyclix-DFT, including grid spacing, number of points for Brillouin zone integration, vacuum in the radial direction, and structural relaxation tolerances (both cell and atom) are chosen such that the computed torsional moduli are numerically  accurate  to within 1\% of their reported value. In terms of the energy, this translates to the  value at the structural and electronic ground state being converged to within $10^{-5}$ Ha/atom, a relatively stringent criterion that is necessary to capture the extremely small energy differences that occur at low values of twist.

\section{Results and discussion} \label{Sec:Results} 
\begin{figure}[htbp!]
        \centering
        \includegraphics[width=0.5\textwidth]{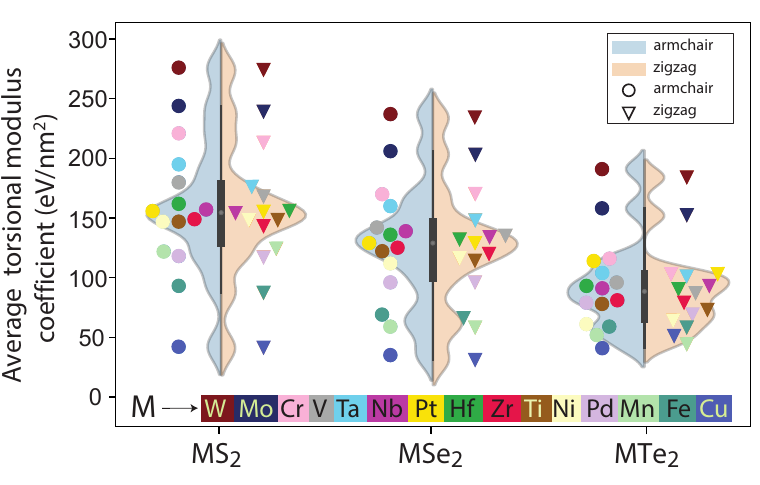}
        \caption{Average torsional modulus coefficient ($\hat{k}$) for the forty-five select armchair and zigzag TMD nanotubes. }
      \label{fig:ViolinTM}
    \end{figure}

As described in the previous section, we utilize symmetry-adapted DFT simulations  to calculate torsional moduli of the forty-five select armchair and zigzag TMD nanotubes. The simulation data for all the results presented here can be found in the Supplementary Material. Observing a power law dependence of the torsional modulus with nanotube diameter $d$, we fit the data to the following relation:
\begin{equation}
K(d) = k d^{\alpha} + K(0) \,,
\end{equation}
where $k$ and $\alpha$ will be henceforth referred to as the torsional modulus coefficient and exponent, respectively. The values so obtained for the different materials  are presented in Table~\ref{tab:Torsion_Modulus_table}. Observing that the exponents are generally close to $\alpha=3$, in order to enable comparison between the different materials that can have nanotubes with significantly different diameters, we also fit the data to the relation:
\begin{equation}
K(d) = \hat{k} d^{3}  + K(0) \,,
\end{equation}
where $\hat{k}$ is referred to as the average torsional modulus coefficient. The results so obtained are presented through violin plots in Figure \ref{fig:ViolinTM}. 

\begin{table*}[htbp]
		\caption{Torsional modulus coefficient ($k$) and exponent ($\alpha$)   for the forty-five select armchair and zigzag TMD nanotubes.}
		\label{tab:Torsion_Modulus_table}
		\centering
		\resizebox{\textwidth}{!}{
			\begingroup
			\renewcommand{\arraystretch}{1.15}
			\begin{tabular}{|c|c|c|c|c|c|c|c|}
				\hline
				\multirow{5}{*}{M}& & \multicolumn{2}{c|}{MS\textsubscript{2} }&\multicolumn{2}{c|}{MSe\textsubscript{2}}&\multicolumn{2}{c|}{MTe\textsubscript{2}}\\
				\cline{3-8}
				&{Diameter}&\multicolumn{2}{c|}{Torsional modulus }&\multicolumn{2}{c|}{Torsional modulus }&\multicolumn{2}{c|}{Torsional modulus }\\
					{}&range&\multicolumn{2}{c|}{coefficient $k$ (eV nm\textsuperscript{$1-\alpha$})}&\multicolumn{2}{c|}{coefficient $k$ (eV nm\textsuperscript{$1-\alpha$})}&\multicolumn{2}{c|}{coefficient $k$ (eV nm\textsuperscript{$1-\alpha$})}\\
				{}&(nm)&\multicolumn{2}{c|}{and exponent $\alpha$}&\multicolumn{2}{c|}{and exponent $\alpha$}&\multicolumn{2}{c|}{and exponent $\alpha$}\\				
				 \cline{3-8} 
			 & &{Armchair}  &{Zigzag}& {Armchair}  &{Zigzag}& {Armchair}  &{Zigzag}\\
			\hline
		{ W}& {2 - 10} &{$267$ $(3.02)$}&{$256$ $(3.04)$} &{$226$ $(3.03)$}&{$230$ $(3.01)$} &{$179$ $(3.04)$}&{$158$ $(3.09)$}\\
		{ Mo}& {2 - 10} &{$232$ $(3.03)$}&{$213$ $(3.07)$} &{$197$ $(3.03)$}&{$175$ $(3.09)$}&{$150$ $(3.03)$}&{$143$ $(3.04)$}\\
		{ Cr} &{6 - 10}&{$194$ $(3.08)$}&{$222$ $(2.98)$} &{$171$ $(3.00)$}&{$178$ $(2.99)$}&{$135$ $(2.95)$}&{$191$ $(2.76)$}\\
		{ V} &{6 - 10}&{$161$ $(3.07)$}&{$170$ $(3.00)$} &{$149$ $(3.00)$}&{$133$ $(3.00)$}&{$98$ $(2.98)$}&{$92$ $(2.97)$}\\
		{ Ta} &{14 - 40}&{$158$ $(3.07)$}&{$205$ $(2.95)$}&{$176$ $(2.97)$}&{$160$ $(2.97)$} &{$165$ $(2.85)$}&{$155$ $(2.86)$}\\
		{ Nb} &{2 - 14}&{$133$ $(3.08)$}&{$181$ $(2.92)$} &{$140$ $(3.00)$}&{$162$ $(2.91)$} &{$90$ $(3.01)$}&{$66$ $(3.17)$}\\
		{ Pt}& {6 - 10} &{$154$ $(3.01)$}&{$156$ $(3.00)$} &{$127$ $(3.01)$}&{$129$ $(3.00)$} &{$133$ $(2.92)$}&{$259$ $(2.53)$}\\
		{ Hf}& {6 - 30} &{$162$ $(3.00)$}&{$165$ $(2.98)$} &{$136$ $(3.00)$}&{$135$ $(2.99)$}&{$93$ $(3.00)$}&{$85$ $(3.02)$}\\
		{ Zr}& {6 - 30} &{$149$ $(3.00)$}&{$160$ $(2.96)$} &{$125$ $(3.00)$}&{$127$ $(2.98)$}&{$98$ $(2.93)$}&{$84$ $(2.98)$}\\
		{ Ti}& {2 - 10} &{$140$ $(3.03)$}&{$153$ $(2.98)$}&{$106$ $(3.08)$}&{$127$ $(2.94)$} &{ $75$ $(3.03)$}&{ $83$ $(2.93)$}\\
		{ Ni}& {6 - 10} &{$147$ $(2.99)$}&{$147$ $(3.00)$} &{$127$ $(2.93)$}&{$120$ $(2.98)$} &{$136$ $(2.63)$}&{$156$ $(2.53)$}\\
		{ Pd}& {6 - 10} &{$114$ $(3.02)$}&{$119$ $(2.99)$} &{$94$ $(3.02)$}&{$100$ $(2.98)$}&{$107$ $(2.85)$}&{$223$ $(2.40)$}\\
		{ Mn}& {6 - 10} &{$108$ $(3.08)$}&{$122$ $(3.00)$} &{ $39$ $(3.23)$}&{ $29$ $(3.38)$} &{ $27$ $(3.40)$}&{ $52$ $(2.99)$}\\
		{ Fe}& {6 - 10} &{$60$ $(3.26)$}&{ $49$ $(3.29)$}&{$102$ $(2.90)$}&{$157$ $(2.71)$}&{ $76$ $(2.87)$}&{ $109$ $(2.66)$}\\
		{ Cu}& {6 - 10} &{ $30$ $(3.19)$}&{ $32$ $(3.13)$} &{ $27$ $(3.14)$}&{ $49$ $(2.76)$}&{ $35$ $(3.20)$}&{ $102$ $(2.35)$}\\
		\hline		
							\end{tabular}
			\endgroup			
		}
	\end{table*}

%%%%%%%%%%%%%%%%%%%%%

%%%%%%%%%%%%%%%%%%%%%

We observe that the torsional modulus coefficients span around an order of magnitude between the different materials, with WS\textsubscript{2} and CuSe\textsubscript{2} having the largest and smallest values, respectively. Notably, even the largest value obtained here is nearly three times smaller than the carbon nanotube (733 eV/nm\textsuperscript{2}) \cite{sharma2021real}, which can be attributed to the extremely strong covalent carbon-carbon bonds. In comparison with literature, where only the values for MoS\textsubscript{2} are available from force field (armchair: 249 eV/nm\textsuperscript{2} and zigzag: 243 eV/nm\textsuperscript{2}) \cite{bucholz2012mechanical}  and tight binding (armchair: 265 eV/nm\textsuperscript{2} and zigzag: 265 eV/nm\textsuperscript{2}) \cite{zhang2010helical} simulations, there is good agreement with the average torsional modulus coefficients reported here (armchair: 244 eV/nm\textsuperscript{2} and zigzag: 239 eV/nm\textsuperscript{2}). Overall, we observe that the torsional moduli values generally follow the trend  MS\textsubscript{2} $>$ MSe\textsubscript{2} $>$ MTe\textsubscript{2}. This can be explained by the metal-chalcogen bond length having the reverse trend, with shorter bonds generally expected to be stronger due to the increase in orbital overlap.

\begin{figure}[htbp]
	\includegraphics[width=0.9\textwidth]{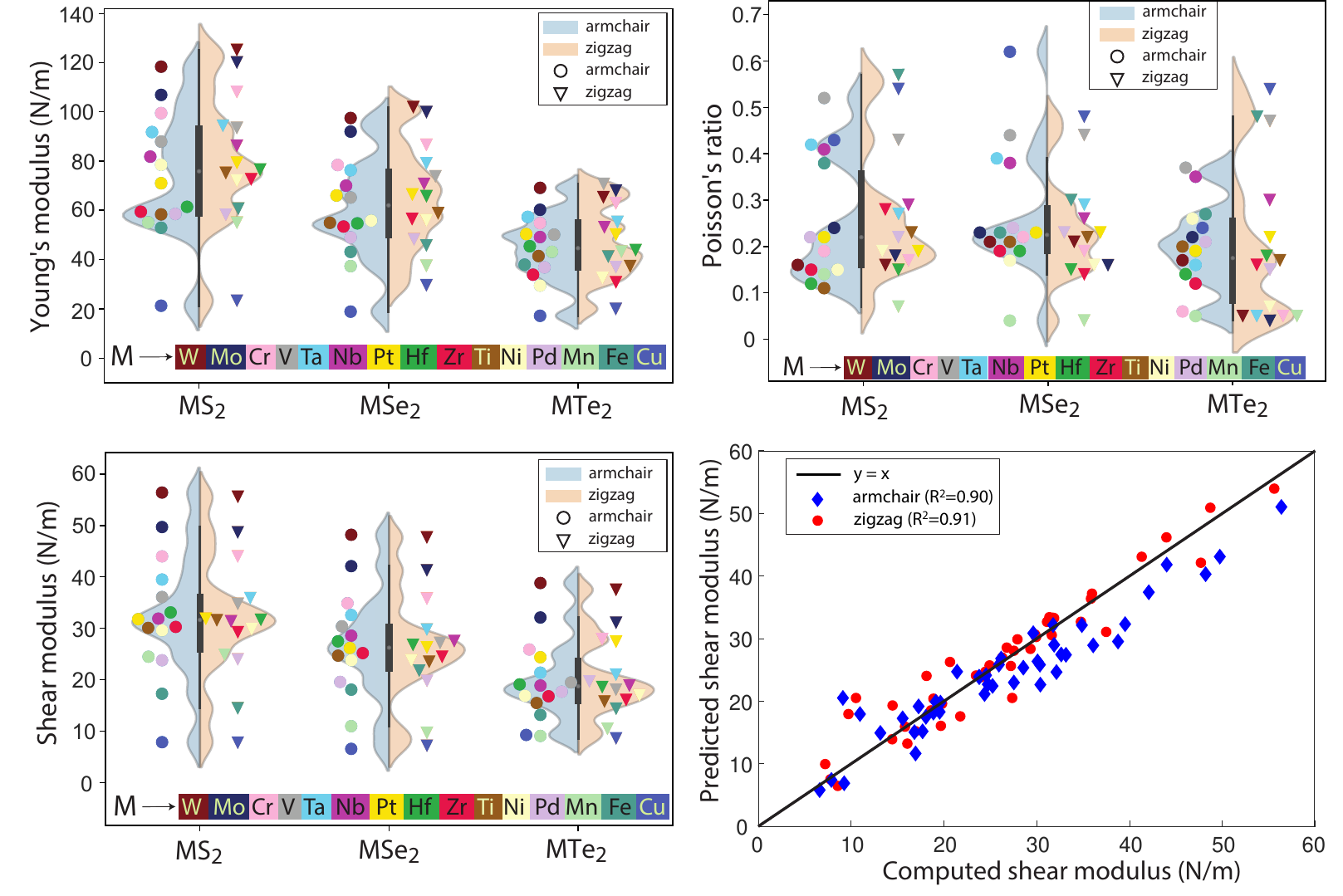}
\caption{ Young's modulus ($E$), shear modulus ($G$), and Poisson's ratio ($\nu$) of the forty-five select armchair and zigzag TMD nanotubes.  The  predicted shear modulus refers to that obtained from the isotropic relation in terms of the Young's modulus and Poisson's ratio. The values of $R^2$ shown in the legend denotes the coefficient of determination for the linear regression. The results correspond to the largest diameter nanotube that has been studied for each material. \vspace{-1mm} }
\label{fig:violin2Iso}
\end{figure}

We also observe from the results in Table~\ref{tab:Torsion_Modulus_table} that the torsional modulus exponents are in the neighborhood of $\alpha=3$, in agreement with the isotropic elastic continuum  model \cite{ugural2003advanced}.  In such an idealization, the shear modulus $G$ can be calculated from the torsional modulus coefficient using the following relation derived from the continuum analysis of a homogeneous isotropic circular tube subject to torsional deformations: 
\begin{equation} \label{Eq:ShearTMC}
G = \frac{\hat{k}}{2\pi} \,.
\end{equation}
The results so obtained are presented in Figure~\ref{fig:violin2Iso}. Note that since there are some noticeable deviations from  $\alpha=3$ (Table~\ref{tab:Torsion_Modulus_table}) --- suggests that the shear modulus changes with diameter --- the results in Figure~\ref{fig:violin2Iso} correspond to the case when $\hat{k}$ is determined from the single data point corresponding to the largest diameter nanotube studied for each material. To  verify their isotropic nature, we  also determine the shear moduli of the nanotubes as predicted by the isotropic relation in terms of the Young's modulus ($E$) and Poisson's ratio ($\nu$): $G = E/2(1+\nu)$, both of which are also calculated using Cyclix-DFT, the results of which are summarized in Figure~\ref{fig:violin2Iso}. It is clear that there is very good agreement between the computed and predicted shear moduli, suggesting that TMD nanotubes can be considered to be elastically isotropic.

We observe from the results in Figure~\ref{fig:violin2Iso} that the Young's moduli follow a similar trend as the torsional modulus coefficients and therefore the shear moduli, which can be again explained by the strength of the metal-chalcogen bond, consistent with results obtained for molybdenum and tungsten TMD monolayers \cite{pub.1009560699, fan2015electronic, pike2019vibrational}. In regards to the Poisson's ratio, we find that the  MnS\textsubscript{2}, MnSe\textsubscript{2}, MnTe\textsubscript{2}, CrTe\textsubscript{2}, WTe\textsubscript{2}, MoTe\textsubscript{2}, TaTe\textsubscript{2}, and NiTe\textsubscript{2} nantotubes have a value near zero. In addition, the CuS\textsubscript{2}, CuSe\textsubscript{2}, CuTe\textsubscript{2}, VS\textsubscript{2} and FeS\textsubscript{2} nanotubes have $\nu$ greater than the isotropic theoretical limit of 0.5, which can be justified by the anistropic nature of these materials --- evidenced by the relatively poor agreement between the predicted and computed shear moduli (Figure~\ref{fig:violin2Iso}) --- where this bound is not applicable \cite{ting2005poisson}. In regards to failure of these materials, it is possible to use Frantsevich's  rule \cite{frantsevich1983elastic} --- materials with $\nu > 0.33$ and $\nu < 0.33$ are expected to be ductile and brittle, respectively --- to predict that M$=$\{Cu, Nb, Fe, Ta and V\} nanotubes are ductile and M$=$\{W, Mo, Cr, Pt, Hf, Zr, Ti, Ni, Pd, Mn\} are brittle. In particular, there is a clear divide between the Poisson's ratio of these two sets, as seen in Figure~\ref{fig:violin2Iso}. Note  that the computed Young's moduli and Poisson's ratio values are in good agreement with those available in literature \cite{zibouche2014electromechanical, zhang2010helical, wang2016strain, bandura2018calculation, ansari2015ab, lorenz2012theoretical}, further confirming the fidelity of the simulations performed here.

\begin{figure}[htbp!]
        \centering
        \includegraphics[width=0.48\textwidth]{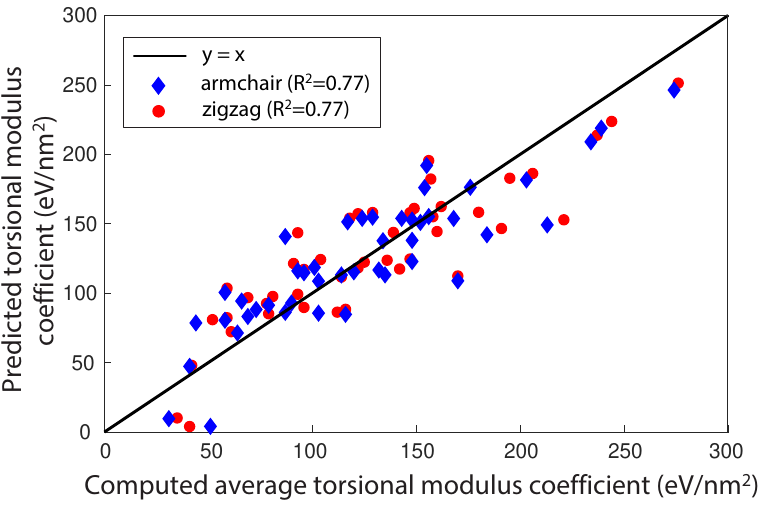}
        \caption{The set of computed average torsional modulus coefficients ($\hat{k}$) and its linear regression with the features being the metal-chalcogen bond length, difference in electronegativity between the metal and chalcogen atoms, and sum of the metal's ionization potential and chalcogen's electron affinity. The values of $R^2$ shown in the legend denotes the coefficient of determination for the linear regression.}
       \label{fig:TMregg}
    \end{figure}

The above results indicate that the torsional moduli of TMD nanotubes are dependent on the nature and strength of the metal-chalcogen bond, which can be expected to depend on the bond length, difference in electronegativity between the  atoms, and sum of their  ionization potential and electron affinity. The first feature mentioned above is used to mainly capture the strength of the bond, and the other two features are used to mainly capture the nature of the bonding \cite{glebko2018electronic, sanderson1986electronegativity, winter2000webelements, morris1956ionization}. Using these three features, we perform a linear regression on the set of average torsional modulus coefficients, the results of which are presented in Figure~\ref{fig:TMregg}. The fit is reasonably good, suggesting that the features chosen here play a significant role in determining the torsional moduli of TMD nanotubes. Note that inclusion of the bond angle as a  feature did not improve the quality of the fit, and therefore has been neglected here.  Also note that though the quality of the fit can be further increased by using higher order polynomial regression, it can possibly lead to overfitting, and is hence not adopted here.  

\section{Concluding remarks} \label{Sec:Conclusions} We have calculated the torsional moduli of forty-five select single-walled TMD nanotubes using \emph{ab initio} DFT simulations. Specifically, we have computed torsional moduli for the armchair and zigzag variants of the chosen TMD nanotubes at practically relevant twists and nanotube diameters, while considering materials that have been synthesized or are likely to be synthesized. We have found that the variation of the torsional moduli values between the different nanotubes follows the trend: MS\textsubscript{2} $>$ MSe\textsubscript{2} $>$ MTe\textsubscript{2}. In addition, we have found that the moduli display a power law dependence on the diameter, with the scaling generally close to cubic,  as predicted by the isotropic  elastic continuum model. In particular,  the shear moduli so determined have been found to be in good agreement with that predicted by the isotropic relation in terms of the Young's modulus and Poisson's ratio, both of which have also been calculated  in this work from  DFT simulations.  Finally, we have developed a linear regression model for the torsional moduli of TMD nanotubes that is based on the nature and characteristics of the metal-chalcogen bond, and  have shown that it is capable of making reasonably accurate predictions. 

In regards to future research, given their significant applications in semiconductor devices, the electromechanical response of TMD nanotubes to torsional deformations presents itself as an interesting topic worthy of pursuit. In addition, given the plethora of multi-walled TMD nanotubes that have been synthesized, studying the effect of torsional deformations on their mechanical and electronic response also presents itself as a worthy subject of investigation.

\section*{Acknowledgements} 
The authors gratefully acknowledge the support of the U.S. National Science Foundation (CAREER-1553212 and MRI-1828187). P.S. acknowledges discussions with Arash Yavari regarding anistropic elasticity.  \vspace{-1mm}

\bibliographystyle{unsrt}
% \bibliography{refer.bib}

\end{document}